\documentstyle[aps,prb,epsfig,floats,twocolumn]{revtex}
\begin{document}

\draft
\flushbottom
\twocolumn[
\hsize\textwidth\columnwidth\hsize\csname @twocolumnfalse\endcsname

\title{Plastic depinning in artificial vortex channels: competition between bulk and boundary nucleation}
\author{T.\ Dr\"ose$^{1}$, R. Besseling$^{2}$, P.\ Kes$^{2}$, and
C.~Morais Smith$^{3}$}
\address{$^1$ I Institut f{\"u}r Theoretische Physik, Universit{\"a}t Hamburg,
D-20355 Hamburg, Germany. \\
$^2$ Kamerlingh Onnes Laboratorium, Leiden University, P.O. Box 9504, 2300 R A Leiden,
the Netherlands. \\
$^3$ Institut de Physique Th\'eorique, P\'erolles, CH-1700 Fribourg,
Switzerland.}

\date{\today}
\maketitle
\tightenlines
\widetext
\advance\leftskip by 57pt
\advance\rightskip by 57pt

\begin{abstract}
We study the depinning transition of a driven chain-like system in the presence
of frustration and quenched disorder. The analysis is motivated by recent
transport experiments on artificial vortex-flow channels in superconducting
thin films. We start with a London description of the vortices and then map
the problem onto a generalized Frenkel-Kontorova model and its continuous
equivalent, the sine-Gordon model. In the absence of disorder, frustration
reduces the depinning threshold in the commensurate phase, which nearly
vanishes in the incommensurate regime. Depinning of the driven frustrated
chain occurs via unstable configurations that are localized at the boundaries
of the sample and evolve into topological defects which move freely through
the entire sample. In the presence of disorder, topological defects can also
be generated in the bulk. Further, disorder leads to pinning of topological
defects. We find that weak disorder effectively reduces the depinning
threshold in the commensurate phase, but increases the threshold in the
incommensurate phase.
\end{abstract}
]
\def\gsim{\mathrel{\raise.4ex\hbox{$>$}\kern-0.8em\lower.7ex\hbox{$\sim$}}}
\def\lsim{\mathrel{\raise.4ex\hbox{$<$}\kern-0.8em\lower.7ex\hbox{$\sim$}}}

\narrowtext

\section{Introduction}
The depinning transition in driven systems has attracted a great
deal of attention recently. The phenomenon can be observed in charge density
wave systems submitted to an
electric field,\cite{Grun} in magnetic bubbles moving under an applied
magnetic field gradient,\cite{Sesh} in current driven Josephson junction
arrays,\cite{Vino,Bale} and in the two-dimensional electron gas in a magnetic
field, which at low densities forms a Wigner crystal, but can move under an
applied voltage.\cite{Bhat} Depinning is
also important to understand tribology and solid friction,\cite{Cule}
surface growth of crystals with quenched bulk or substrate disorder, and the
dynamics of
domain walls in incommensurate solids.\cite{Pokr} A very prominent example
of driven systems displaying a depinning transition are vortices in type-II
superconductors.\cite{Blat}

The $H-T$ phase diagram of type-II superconductors displays regions of
both elastic and plastic flow.\cite{Bhat,Bhat2}
The interplay between the elasticity of the
vortex lattice and the impurities present in the substrate leads to a rich
phenomenology with many static and dynamic phases. A crucial question in both
the dynamics and statics is whether - in addition to thermal fluctuations -
quenched disorder produces topological defects in the periodic structure.
Whereas in the absence of topological defects it is sufficient to consider only
elastic deformations with depinning causing elastic flow, the dynamics will
be governed by plastic flow, if topological defects exist.
One expects plastic motion to become important for either strong disorder,
high temperature, or near the depinning transition in low dimensions.
In these cases, depinning
is observed to occur through ``plastic channels'' between pinned regions.
This type of plastic flow has been found in numerical simulations of a
two-dimensional thin film geometry.\cite{Jens1,Jens2,Jens3} Above the
threshold, the filamentary channels become both denser and broader when the
driving force is increased. Measurements
of the differential resistance of MoGe films display abrupt steps, which could
be interpreted in terms of plastic depinning.\cite{Hell}
Other experimental observations
that have been attributed to plastic flow are the peak effect,\cite{Word,Berg}
unsual broadband noise,\cite{Marl} and steps in the $I-V$ curve.\cite{Dann}

In order to study the plastic depinning of vortices, artificial easy-flow
channels have been manufactured.\cite{Pruy,Theu} The samples are typically
made of type-II superconducting thin films, which consist of a fairly strong
pinning layer on top of a weakly pinning base layer. The artificial vortex
flow channels are fabricated by etching away stripes of the stronger pinning
top layer. In the Shubnikov phase, vortices penetrate the sample. Applying
a current perpendicular to the channel direction, the resulting Lorentz force
in the direction of the channel drives the vortices. Since the influence of
point-like material defects in the weak pinning channel is negligible,
channel vortices are mainly pinned indirectly via the interaction with the
stronger pinned vortices in the channel environment. Above a threshold force,
plastic depinning of the vortices inside the channel takes place. In contrast
to natural channels, the depinning threshold force displays interesting
oscillations when the externally applied magnetic field is varied. The
position of the threshold force maxima hint at commensurability effects
between the vortex lattice in the environment and in the channel.\cite{Bess}
The magnitude of the depinning force minima and maxima indicate that lattice
distortions produced by quenched disorder in the pinned channel environment
is relevant in these samples.\cite{Rut}

In this work, we present theoretical studies of the depinning and dynamical
behavior of vortices in artificial easy-flow channels. Frustration and disorder
are essential ingredients for determining the depinning transition in this
driven system. Motivated by recent experiments on vortex-flow
channels, we develop a model which takes into account inhomogeneities, such as
the sample boundaries and disorder. We find that depinning occurs due to boundary
or bulk nucleation, respectively, in the absence or presence of disorder. Further,
we clarify the role of disorder and frustration and compare our results to the
available experimental data.

First, we study this problem considering a perfectly ordered vortex-lattice
in the environment. Starting from a London description of
vortices inside and outside the channel, we derive the coefficients of
a generalized Frenkel-Kontorova model.\cite{Frenk}
The force-velocity curve of this simple model displays a
drastic decrease of the threshold pinning force when topological defects
enter the system via the system boundaries. However, the experimental data
show a smooth variation of the pinning force as a function of the magnetic
field. The behavior suggests that disorder in the channel edges may lead
to smoothening of the transition from the topologically ordered to the
defective state. In order to investigate the transition in presence of disorder,
we take into account that the vortex lattice in the channel environment is
distorted due to impurities in the superconducting substrate. To determine
the characteristic current-voltage or force-velocity curves, the problem is
most conveniently tackled with numerical molecular dynamics simulations. This
method enables us also to nicely visualize how depinning occurs in the
topologically ordered and defective regimes. It turns out that systems that
are topologically ordered in the static phase depin at the boundary or by
generating mobile defect droplets at weak spots in the random channel potential.
In the defective static phase, the pre-existing topological defects are
pinned by the random potential. Then, the depinning transition occurs via
a release of these defects at the threshold.

The paper is organized as follows: In Section II, we discuss the experiments
that motivated the theoretical studies; in Section III we introduce the model
and present the results concerning the depinning transition.
Finally, the discussion and conclusions are presented in Section IV.

\section{Experimental Motivation}\label{sec:motivation}
\begin{figure}[t]
  \begin{center}
    \epsfig{file=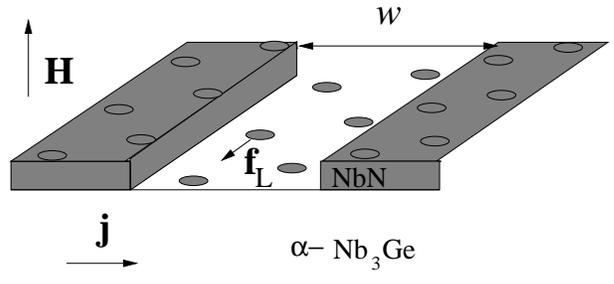,width=8cm}
  \end{center}
  \vspace*{-0.2cm}
  \caption[]{\label{artchan}
Schematic plot of the artificial flow channel geometry. The channels
are manufactured by etching away the strongly pinning top layer material
(grey area) in stripes of width $w $. Applying a magnetic field $ {\bf H} $
generates vortices. In the strong pinning area the vortices are static.
Vortices in the weakly pinning channel are mainly pinned due to the interaction
with the strongly pinned ones in the channel edges (grey area).
In presence of a current density $ {\bf j} $ a Lorentz force
$ {\bf f}_{L} $ acts on the vortices. At the depinning transition the
channel vortices start to flow whereas the strongly pinned vortices in the
channel environment remain static.}
\end{figure}

\begin{figure}
\begin{center}
\epsfig{file=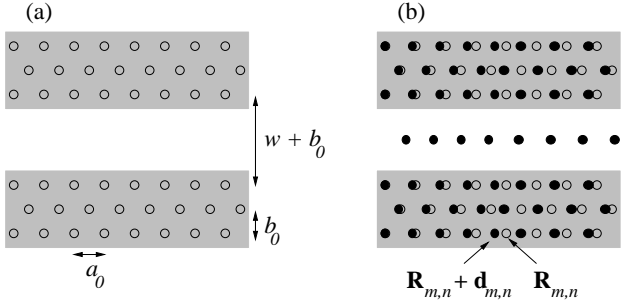, height=4cm}
\vspace{.1cm}
\caption{(a) Easy flow channel environment (shaded) in the absence of disorder.
Vortex positions in the environment are marked by open circles.
(b) Immobile vortices (dots) in the environment in the presence of weak
disorder and one row of mobile vortices in the flow channel. }
\label{channel1}
\end{center}
\end{figure}

Let us start with a short overview of the experimental status of art.
In order to obtain a more detailed understanding of the depinning process
in disordered superconducting films, experimentalists fabricated
{\em artificial} vortex channels.\cite{Pruy,Theu}
A sketch of a typical device is presented in Fig.\ \ref{artchan}.
The devices are manufactured by first absorbing a thin layer of NbN (strong
pinning material) on top of a Nb$_3$Ge film (weak pinning) and then
etching away a few hundred channels of width $ w $ from the top layer.
In the strongly pinned region, the vortex-lattice spacing is
$ a_{0} = (4/3)^{1/4}(\Phi_{0}/B)^{1/2}$, such that $a_0 b_0 = \Phi_0 / B$,
with $\Phi_0$ the flux quantum and $b_0 = (\sqrt{3}/2) a_0$, see Fig.\ 2a.
In order to measure the effect of a single vortex row in the channel, the
channel width $w$ is chosen to be of the order of $ a_0 $.
Therefore, the vortex-lattice spacing in the channel is $a = \Phi_0 / w B$.
Depending on the mismatch parameter $\chi = (a/a_0) - 1 = (b_0 / w) - 1$,
a commensurate-incommensurate transition can be observed. In the experiments
described here,
$ a_{0} \sim w \sim 100 $ nm  and the distance between the channels is
$ 10~ \mu$m. Due to the material steps at the channel edges,
screening currents flow along the surface which interact with the vortices
in the channels. As a consequence, we can expect the first rows of strongly
pinned vortices near the channel edges to be lined up at fairly straight
along the channel. Therefore, we can assume that the transversal shifts are
negligible.
On the other hand, longitudinal displacements
may occur due to the presence of impurities
in the superconducting material [see Fig.\ \ref{channel1}(b)].
Here we will concentrate on the longitudinal dynamics of the vortices
within the channel.

\begin{figure}[t]
  \begin{center}
    \epsfig{file=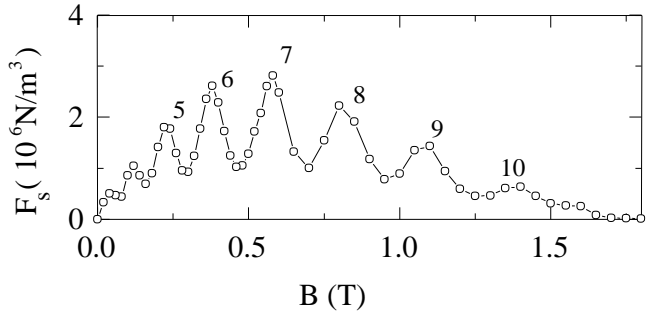,width=8.5cm}
  \end{center}
  \vspace*{-0.2cm}
  \caption[]{\label{chanexp}
The experimentally obtained critical shear force density $ F_{s} $
versus magnetic field $B$ for a channel sample with $ w \approx
290~$nm at a temperature $T=1.7$ K \cite{rutlong}.
            }
\end{figure}

In the channel, pinning due to impurities is very weak. Below the depinning
threshold, vortices in the channel are localized due to the interaction
with the vortices in the static environment.
In the presence of a current density ${\bf j}$, a Lorentz force
$ {\bf f}_{L} $ acts on the vortices. As long as the vortices are pinned,
the supercurrent is flowing dissipation-less through the sample.
At a threshold force $  f_{c} $ the channel
vortices depin and start to move. The vortex motion causes dissipation of
energy and  a voltage drop occurs across the sample
boundaries. Hence, a well-defined critical current density $ j_{c} $
can be determined by measuring the current-voltage characteristics of
the sample. Fig.\ \ref{chanexp} shows shear force density data deduced from
current-voltage measurements.\cite{rutlong}
The oscillations in the shear force density $ F_{s} $
as a function of $ B $
are related to the (in)commensurability of the vortex lattice $b_0$
with the channel width $ w $. Naively, we would expect the
maxima of $ F_{s}$ to occur at integer values of
the ratio $ w / b_{0} $, where the vortex lattice spacing inside the channel
is commensurate with the width of the channel.
$ F_{s}$ should then be reduced for non-integer $ w/b_{0}$.
This could qualitatively be
explained by the development of misfit dislocations along the channel
edges.\cite{Pruy,Theu} Although this intuitive picture holds for the case
of weak (or zero) disorder, these features change drastically in the limit
of a strongly disordered environment. Indeed, as it will be explained
in Ref.\ [\cite{rutlong}], in the latter limit, the minima and not the
maxima of $F_s$ correspond to an integer number of vortex-rows in the
channel. In this paper, however, we restrict ourselves to the limit of
weakly disorderd environment and investigate the competition between
boundary and bulk nucleation for activating the depinning process.

\section{A model for artificial vortex channels}
Starting from a London description of vortices in a type-II
superconductor, we now derive a simplified 1d model for
vortices in an artificial easy-flow channel.
\subsection{Channel with a perfect vortex lattice in the environment}
Let us first consider a row of straight vortices that are aligned
along the $x-$axis at positions $ m a_{0} {\bf e}_{x} $,
where $m$ is an integer, $a_{0}$ the
distance between adjacent vortices and $ {\bf e}_{x} $ the unit vector
in the $x-$direction. In a type-II superconductor within London theory
the interaction energy between two straight vortices of length $l$ at
$ {\bf r}$ and $ {\bf 0}$ is $ U({\bf r} ) = U_{0} K_{0}(r/\lambda)  $,
where $ U_{0} =  \Phi_{0}^{2} l/(8 \pi^{2} \lambda^{2})$.
Here $ \lambda $ is the penetration depth, $\Phi_{0}=hc/2e $ is the flux
quantum, and $l \gg \lambda $ the sample thickness.
Then the potential energy $ E^{o}_{n}({\bf r})$
that is felt by a test vortex at
position $ {\bf r} = (x,y)$ interacting with a row of vortices
placed at $ n b_{0} {\bf e}_{y}  $ is
\begin{equation}
E^{o}_{n}({\bf r})=\sum_{m} U({\bf r}- {\bf R}_{m,n} )
\end{equation}
where we introduced the lattice vectors
$ {\bf R}_{m,n} =
 ( [m+ n/2] a_{0}, n b_{0})$
 of a hexagonal lattice.
Fourier transforming and using the Poisson sum rule the latter can be recast
into
\begin{equation} \label{ordrow}
E^{o}_{n}({\bf r})= \sum_{\nu} \frac{\cos[q_{\nu}(x-na_{0}/2)] }{2 \pi a_{0}}
\int_{-\infty}^{+\infty} dk_{y} {\tilde U} e^{i k_{y} (y- n b_{0})}
\end{equation}
where $ q_{\nu} =2 \pi \nu/a_{0}$ with integer $\nu$
and $\tilde U$ is the Fourier transform of the potential, which is given by
\begin{equation}
{\tilde U(q_{\nu},k_{y})}=\frac{2 \pi U_{0} }{ q_{\nu}^{2} + k_{y}^{2}
 + \lambda^{-2}}.
\end{equation}
Performing the integral  over $ k_{y}$ leads to
\begin{equation}
E^{o}_{n}({\bf r})=
\sum_{\nu} B_{\nu}(|y- n b_{0}|)~ \cos[q_{\nu}(x-na_{0}/2)],
\end{equation}
where
\begin{equation}
B_{\nu}(y)= \frac{ \pi U_{0} }
     { a_{0}q_{\nu} ' }~ {\rm e}^{-  q_{\nu}' y  }
\end{equation}
and  $ q_{\nu}' = \sqrt{q_{\nu}^{2}+\lambda^{-2}}$.

We construct an easy flow channel by building a two dimensional vortex
lattice, but leaving a region of width $ w$
along the $x-$axis empty (see Fig.\ \ref{channel1}a).

If we consider a hexagonal vortex lattice in the channel environment,
the potential in the channel is
\begin{eqnarray}
E_{oc}({\bf r}) & = & \sum_{n = 1}^{\infty}
\left[
 E^{o}_{n} \left( {\bf r}- {\bf b}/2 \right)
+E^{o}_{-n} \left( {\bf r}+ {\bf b}/2 \right)
\right],
\end{eqnarray}
where  ${\bf b}= (w-b_{0}) {\bf e}_{y} $ .
Summing over the vortex rows $ n $ one finds
\begin{equation}
E_{oc}({\bf r}) =
 \sum_{\nu} A_{\nu}(y)  \cos{(q_{\nu}x)},
\end{equation}
the Fourier coefficients $ A_{\nu}(y)  $ read
\begin{equation}
A_{\nu}(y) =
\frac{2  \cosh(q_{\nu}'y)
 B_{\nu} \left( \frac{w-b_{0}}{2} \right)}
{ (-1)^{\nu} {\rm e}^{  q_{\nu}' b_{0}  } -1  }.
\end{equation}
Since the magnetic inductions $ B $ used in the experiments
cover the entire range
up to the upper critical one  $ B_{c2}$, a more general expression
for $ A_{\nu} $ than the London limit that has been discussed until now
is needed.
However, the theory presented here can easily be extended to larger magnetic
inductions. First one takes into account the finite diameter of the vortex
core which is of the order $ \xi$ in the vortex-vortex interaction potential,
$ U \to U_{0}[K_{0}(r/\lambda)-K_{0}(r/\xi)] $. Second, one replaces
$ \lambda \to \lambda'= \lambda/(1-B/B_{c2})^{1/2} $ and
$ \xi \to \xi'=\xi/(2-2B/B_{c2})^{1/2} $ to take into account the reduction of
the superconducting order parameter at large magnetic  fields.\cite{Bran}
We then obtain
\begin{eqnarray} \nonumber
A_{\nu}(y) &=& \frac{ \pi U_{0}(1-B/B_{c2}) } { a_{0}}
\left[
\frac{2  \cosh(q_{\nu}'y)
 {\rm e}^{-  q_{\nu}' y(w-b_{0})/2} }
{q_{\nu}' \left[ (-1)^{\nu} {\rm e}^{  q_{\nu}' b_{0}  } -1 \right]  } \right. \\
&-& \left.
\frac{2  \cosh(q_{\nu}''y)
 {\rm e}^{-  q_{\nu}'' y(w-b_{0})/2} }
{q_{\nu}'' \left[ (-1)^{\nu} {\rm e}^{  q_{\nu}'' b_{0}  } -1 \right]  }
\right],
\end{eqnarray}
where now $ q_{\nu}' = \sqrt{q_{\nu}^{2}+(\lambda')^{-2}}$ and
$  q_{\nu}'' = \sqrt{q_{\nu}^{2}+(\xi')^{-2}}$.

\subsection{Equation of motion} \label{sec:FKM}
The overdamped dynamics of a vortex with index $m$ at position
$ {\bf r}_{m} $ inside the channel is described by
\begin{equation} \label{equofmot}
{\eta} {\dot {\bf r}_{m}} =
f \cdot {\bf e}_{x} - \nabla E_{oc}({\bf r}_{m})
- \sum_{n \not = m} \nabla U({\bf r}_{m}-{\bf r}_{n}),
\end{equation}
where $ f = j \Phi_{0}  /c $ is the magnitude of the
Lorentz force that drives the vortices
in presence of a current density  $j$.
The viscous drag coefficient $ {\eta} $
is related to the flux flow resistivity $ \rho_{ff}$ by
${\eta} = B \Phi_{0} /[c^2 \rho_{ff}(B)]$.
The sum is taken over the positions $n$ of all other vortices inside
the channel.

In the simplest case the channel width is $ w \approx b_{0}  $ such that
only a single vortex row is inside the channel.
Then the motion of the mobile vortices in the
$y-$direction is essentially guided by the channel potential, whereas the
interaction between vortices in the channel does not contribute significantly
to the motion in the $y-$direction,
$ {\eta} {\dot y}_{m} \approx -\partial_{y} E_{oc}({\bf r})$.
However, the motion in the x-direction is determined by both the interactions
between mobile vortices and the gradient of the channel potential.
To simplify matters, we neglect the motion in the $y-$direction, $y_{m}=0$,
such that the equation of motion can be simplified to a one-dimensional one.
Further, we restrict our considerations to  $ \lambda \gsim a_{0}$ as in
a typical experiment. Then, the amplitudes $ A_{\nu}(0) $ fall off
exponentially fast, $ A_{\nu}(0) \sim U_{0} \exp(- \pi \sqrt{3} \nu),$ and the
approximation to consider only the first harmonic $ q=q_{1}=2 \pi/a_{0}$
of the channel potential is a good one. Introducing $\mu = 2 q |A_{1}(0)| $,
and restricting the interaction between vortices in the channel to
next-neighbor spring-like forces, the equation for the overdamped
longitudinal motion reads
\begin{eqnarray} \label{FKM}
{\eta} {\dot x}_{m} & =& -\frac{\partial V}{\partial x_{m} },
\end{eqnarray}
where the potential energy of the vortices in the channel is given by
\begin{eqnarray} \nonumber
 V &=& \sum_{m} \left\{
 \frac{\mu}{q} \left[1- \cos \left( q x_{m}\right) \right] - f x_{m} \right. \\
&+& \left. \sum_{n} \frac{\kappa_{n}}{2} (x_{m+n} - x_{m} -na)^{2} \right\}.
\label{UFKM}
\end{eqnarray}
It has the form of a generalized Frenkel-Kontorova model.\cite{Frenk}
The interactions between vortices inside the channel
are approximated by Hookian springs with spring constants
$ \kappa_{n} = (U_{0}/\lambda^{2}) K''_{0}(na/\lambda)$, where the double prime
denotes the second derivative.
Notice that the displacements $x$ are measured with respect to the vortex-lattice
preferred position $a$ in the channel.

The Frenkel-Kontorova model has been intensively
studied close to equilibrium, $ f \sim 0$, see Ref.\ [\cite{Brau}].
The reduced dimensionless elasticity
\begin{equation} \label{g}
g = \frac{1}{q \mu}  \sum_{n=1}^{\infty} n^{2} \kappa_{n}
\approx \frac{ e^{\pi\sqrt{3}} }{8 \pi} \frac{\lambda}{a} \gg 1
\end{equation}
together with the winding number ${\Omega} = a/a_{0}$
crucially determine the behavior of the system.
For rational $ {\Omega} $
the vortex chain is commensurate with the periodic channel potential whereas
for irrational $ {\Omega} $ it is in an incommensurate state.
The commensurate-incommensurate transition is a continuous transition
that occurs  at finite mismatches $\chi$ (recall that $\chi = (a - a_{0})/a_{0}$),
since the creation of discommensurations costs energy.

If the discreteness of the chain is relevant,
Peierls-Nabarro barriers may exist.
The Peierls-Nabarro is the energy barrier that has to be overcome for a
translation, $ x_{m} \to x_{m+1} $.
Whereas this barrier is always finite for rational
${\Omega}$ , it may vanish in the incommensurate state:
if $ g $ is lower than a critical value, $ g_{c}(a/a_{0})$, the incommensurate
state is pinned,
however, for $ g > g_{c}(a/a_{0})$, the Aubry transition to a sliding state
with truly vanishing critical force takes place.\cite{Aubr}
For pinned defective configurations the Peierls-Nabarro barrier,
which determines the corresponding pinning force $ f_{PN} $,
depends on $g$. Large
$ g \gg 1 $ imply that an isolated defect having a size
$ \sim l_{d} $ extends over several lattice
constants. Then, the Peierls-Nabarro barrier is nearly vanishing
\cite{Brau} and the pinning force is
\begin{equation}
f_{PN} \approx 64 \pi^{2} g \mu  \exp(-\pi^{2} \sqrt{g}).
\end{equation}
Since $ f_{PN} \ll 10^{-4} \mu $ we neglect it in the following.

Since we are most interested in the regime $ g  \gg 1$, it is
convenient to study the model in the continuum
limit. Further, we take into account the finite length of the system and
consider open boundary conditions.
Rewriting Eq.\ (\ref{UFKM}) in terms of the displacements
of the vortices from the environmental lattice positions, $ u_{m} = x_{m} - m a_{0} $
and substituting $ m a_{0} \to x$, $ u_{m} \to u(x)$,
$ (u_{m+n} - u_{m})/(na_{0}) \to \partial_{x} u(x)$,
$ \sum_{m} \to \int dx/a_{0}$, we obtain the continuum equation of
motion
\begin{eqnarray}\nonumber
\eta {\dot u}(x) &=& - a_{0} \frac{\delta V[u]}{\delta u(x)}
= {\tilde \kappa}
\frac{\partial^{2} u(x)}{\partial x^{2}}  -
 \mu \sin \left[ q u(x) \right] \\
&+& f + \chi {\tilde \kappa}  [\delta(x-L)-\delta(x)],\label{CFKM}
\end{eqnarray}
where we have introduced ${\tilde \kappa}= 2 \pi a_{0} \mu g$
and the last term is an effective surface force
that arises at the open sample boundaries in the presence
of frustration ($L$ is the channel length).
The energy functional reads
\begin{displaymath}
 V[u] =  \int
\frac{dx}{a_0} \left\{
\frac{{\tilde \kappa}}{2}
\left(\frac{\partial u}{\partial x} - \chi \right)^{2} +
 \frac{\mu}{q} \left[1- \cos \left( q u \right) \right]
- f u
\right\},
\end{displaymath}
where the integral runs from $0$ to $L$.
It can be decomposed into
\begin{equation}
 V = V_{SG} + V_{\chi} + V_{0},
\end{equation}
where
\begin{displaymath}
 V_{SG}[u] =  \int
\frac{dx}{a_0} \left\{
\frac{{\tilde \kappa}}{2}
\left(\frac{\partial u}{\partial x} \right)^{2} +
 \frac{\mu}{q} \left[1- \cos \left( q u \right) \right]
- f u
\right\}
\end{displaymath}
is the energy functional of the sine-Gordon model,
\begin{equation} \label{frustrationenergy}
 V_{\chi} [u] = - \frac{\chi {\tilde \kappa}}{a_0}
 \left[u(L)-u(0) \right]
\end{equation}
is the frustration energy due to the mismatch $ \chi $ determined
by the values of the displacement field $ u(0)$ and $ u(L)$ at the boundaries,
and $ V_{0} $ is an irrelevant offset that is omitted in the following.

\subsection{Commensurate-Incommensurate Transition} \label{CIT}
We now present a short review of the commensurate-incommensurate transition.
Then, we will extend the picture to discuss the role of
edge barriers for defects
in finite systems, which is crucial to understand how discommensurations
penetrate a sample in the absence of thermal or quantum fluctuations.

The extrema of $ V[u] $ are found solving the variational problem
$ \delta V/ \delta u =0$. In the absence of frustration, $ \chi=0$, and for
$f = 0$, uniform static solutions exist. They are the stable, $ u_{s,n} = a_{0} n$,
unstable, $ u_{u,n} = (2n+1) a_{0} /2 $, and kinked
\begin{equation}
 u_{k,n}(x;x_{c}) =  u_{s,n} +\frac{4}{q}
              \arctan \left[ \exp\left( \frac{x-x_{c}}{l_{d}}\right) \right],
\end{equation}
solutions of the sine-Gordon model at zero driving force. Here,
$ l_{d} =  a _{0} \sqrt{g} $ is the width of the kink.
The corresponding anti-kink solution reads
$  u_{a,n}(x;x_{c}) =  u_{k,n}(-x;-x_{c}) $.
The commensurate-incommensurate transition is a transition that occurs in
equilibrium ($f=0$) when it becomes energetically favorable
to have a finite density of discommensurations in the system.
Neglecting effects of system boundaries, the
mismatch $ \chi_{ci} $ at which the commensurate-incommensurate transition takes
place is then quickly
found by comparing the energy  of the kinked solution with the
energy  of the stable homogeneous
one $ V[u_{k,n}]-V[u_{s,n}]= (4/\pi)({\tilde \kappa} \mu / q)^{1/2} -
{\tilde \kappa} \chi =0, $
\begin{equation}
 \chi_{ci} = \frac{4}{\pi} \left(\frac{\mu}{q {\tilde \kappa}} \right)^{1/2}
           = \frac{2}{\pi^{2} \sqrt{g}}.
\end{equation}
Though at the commensurate-incommensurate transition $ u_{k,n}$ or $ u_{a,n} $ have the
same energy as $ u_{s,n}$, a barrier has to be overcome in order to make the
transition from $ u_{s,n}\to u_{k,n}$ or $u_{s,n} \to u_{a,n} $, see upper curve
in Fig.\ 4. In the absence of fluctuations, as considered here, discommensurations
can only enter the system when this barrier vanishes (lower curve in Fig.\ 4).
In general, the barrier vanishes
at a sufficiently strong frustration or driving force.
We therefore define a threshold frustration $ \chi_{c}(f)$,
which is a function of the driving force $f$. Finite driving forces $ f>0$
are considered in Section \ref{subsec:depinning}, where depinning is studied.
In the following we investigate how (anti-)kinks enter the system
at equilibrium $ f=0$ and determine the zero-force threshold $\chi_{c}(0)$.

Let us first determine the energy that is needed to deform a uniform state
into a kinked one. We note that due to frustration kinks can only be created
spontaneously at the system boundaries, $x=0$ or $x=L$.
Of course, in the presence of thermal fluctuations, quantum fluctuations, or
quenched disorder, deep in the bulk kinks can in principle be
created in the form kink-anti-kink pairs. However, in the absence of
fluctuations as considered here this is not possible. The reason is that
kink-anti-kink pairs cannot gain
frustration energy for a spontaneous kink emergence,
since $ u(0)=u(L)=u_{s,n} $ and hence $ V_{\chi}=0$.
Thus, (anti-)kinks  can only
enter the system at $x=0$ and $ x=L $.
In the following we discuss the penetration of an anti-kink at $x=0$,
having in mind that the same holds for $x=L$ and for kinks.
Note that in a finite system with boundaries at $ x=0$ and $x=L$
one has $\lim_{x_{c} \to - \infty} u_{a,n}(0;x_{c}) = u_{s,n}$.
Hence, a deformation
from $ u_{s,n} $ to $ u_{a,n} $ can be achieved by pushing an
anti-kink centered at $ x_{c} $ from $ x_{c} = -\infty $ to $ x_{c} > 0 $.
The energy of the anti-kink solution relative to the uniform
one as a function of center coordinate $ x_{c} $ is then
\begin{eqnarray}
\Delta V(x_{c}) &=& V[u_{a,n}(x;x_{c})]-V[u_{s,n}(x)] \nonumber  \\
&=& a_{0}^{-1} {\tilde \kappa} \left\{
\chi  [u_{a,n}(0;x_{c})- u_{a,n}(L;x_{c})]  \right.  \nonumber \\
&+& \left.   \int_{0}^{L} dx
\left[ \frac{\partial u_{a,n}(x;x_{c})}{\partial x} \right]^{2}
\right\}
\end{eqnarray}
%%%%%%%%%%%%%%%%%%%%%%%%%%%%%%%%%%%%%%%%%%%%%%%%%%%%%%%%%%%%%%%%%
\begin{figure}[t]
\begin{center}
  \epsfig{file=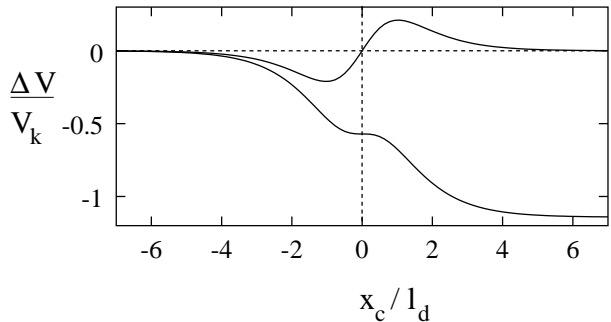,width=8cm}
\end{center}
\vspace*{-0.2cm}
\caption[]{\label{kinkpot}
Potential energy of an (anti-)kink $ \Delta V $ in units of
$ V_{k} = {\tilde \kappa} \chi_{ci}/2$ as a function of
the (anti-)kink center $ x_{c} $. Displayed are values
close to the system boundary at $ x=0 $ for
frustration parameters $ |\chi|=\chi_{ci} $ (upper curve) and
$ |\chi| = \chi_{c}(0) $ (lower curve). At the classical commensurate-incommensurate transition, where
$ |\chi|=\chi_{ci}$, an entry barrier has to be overcome with the
help of fluctuations to make a
transition from $ u_{s,n} $ to $ u_{k,n} $ or $ u_{a,n} $. The entry barrier
vanishes at $ |\chi|=\chi_{c}(0) $ where (anti-)kinks can penetrate the
system.          }
\end{figure}
%%%%%%%%%%%%%%%%%%%%%%%%%%%%%%%%%%%%%%%%%%%%%%%%%%%%%%%%%%%%%%%%%
Since the tail of the kink falls off exponentially, we can neglect the
influence of the boundary at $ x=L \gg l_{d} $.
For simplicity, we then consider a
semi-infinite system, $ L \to \infty$, and obtain
\begin{eqnarray} \nonumber
\Delta V (x_{c}) &=& \frac{{\tilde \kappa} \chi_{ci}}{2} \left\{
\frac{4}{\pi}\frac{\chi}{\chi_{ci}}
\arctan \left[ \exp \left( \frac{x_{c}}{l_{d}} \right) \right]
+1 \right. \\ &+& \left. \tanh \left( \frac{x_{c}}{l_{d}} \right)  \right \}.
\end{eqnarray}
Minimizing $ \Delta V $ with respect to $ x_{c} $ we find a minimum at
\begin{equation}
x_{c,1}=l_{d} \ln
\left(- \frac{\pi \chi_{ci}}{2 \chi}-
\sqrt{\frac{\pi^{2} \chi_{ci}^{2}}{4 \chi^{2}}-1} \right)
\end{equation}
and a maximum at
\begin{equation}
x_{c,2}=l_{d} \ln
\left(-\frac{\pi \chi_{ci}}{2 \chi}+
\sqrt{\frac{\pi^{2} \chi_{ci}^{2}}{4 \chi^{2}}-1} \right).
\end{equation}
At the frustration
\begin{equation} \label{chi_c}
\chi= -\chi_{c}(0)= -\frac{\pi}{2} \chi_{ci}= - \frac{1}{\pi \sqrt{g}},
\end{equation}
the minimum and the maximum merge into a saddle point at $ x_{c}=0$ where
the entry barrier vanishes and an anti-kink flows freely  into the system.
It is interesting to note that for $ \chi < 0 $ the minimum
of the anti-kink energy relative to the uniform solution is always negative,
$ \Delta V(x_{c,1}) < 0 $. This means that in a frustrated system
the uniform solution $ u_{s,n} $ is unstable in the presence of
a boundary. Instead, the stable solution is a virtual
anti-kink with a center $ x_{c} $ localized outside the system at
$x_{c,1} <0$. At the boundary, the chain thus tries to adapt
optimally to the frustration to reduce its energy.
For $ \chi=-\chi_{c}(0) $
one finds $ x_{c} = 0$, which means that half of the kink is already
inside the system and that it can gain more energy by sliding towards the
center of the system.
The scenario is the same for an anti-kink entering the system at
$ x=L $. For a kink the description given above is identical except that
$ \chi > 0$. The kink entry barrier vanishes at $ \chi=\chi_{c}(0) $.

The picture of the commensurate-incommensurate transition is thus drastically modified
in the presence of system boundaries when there are no physical mechanisms
like thermal or quantum fluctuations that are needed to cross the
edge-barrier. In fact, since in the absence of fluctuations
a system with boundaries remains commensurate for $| \chi |< \chi_{c}(0)$,
we identify the threshold at $ |\chi|=\chi_{c}(0) $
with the commensurate-incommensurate transition of a finite, purely mechanical system.

\subsection{Depinning in the presence of boundaries}
\label{subsec:depinning}
In the following we investigate how the chain inside the channel actually
depins in presence of a driving force, $f>0$.
In the simplest case, for $  w=b_{0}  $,
we have a commensurate state without frustration,
$ a=a_{0}$ and $ \chi=0$. The chain locks perfectly to the
potential and the threshold depinning force is
\begin{equation}
 f_{c}(0)=\mu.
\end{equation}

If the system is frustrated, $ b_{0} \not= w$,
depinning occurs via mobile discommensurations which in the sine-Gordon
model are represented by kinks or anti-kinks.
As in the equilibrium case, for finite driving forces
discommensurations enter the system when
their entry barrier vanishes.
Whether they are mobile or not depends
on further barriers that may exist in the bulk.

In the continuous limit as discussed here, the defective state is not pinned.
Hence, in the continuum model the threshold  $ \chi_{c}(0)$
indicates the change from a
static equilibrium ground state ($ u_{s,n} $) to a mobile one ($ u_{k,n} $ or
$ u_{a,n} $).

Note that an exit barrier exists for a {\em single}
kink at the other boundary (imagine the mirror image of the entry barrier
as shown in Fig.\ \ref{kinkpot} at the other end of the system).
However, the exit barrier becomes irrelevant in the presence of further
kinks. This can be easily understood by the following argument. Suppose
a kink enters the system, freely flows to the other end, and then becomes
trapped by the exit barrier. Then a second kink follows and interacts with
the first one. If it would move ``adiabatically'' it would become trapped
by the interaction with the first kink, which mediates the pinning force.
However, the first kink would experience the interaction of the second kink,
too. The resulting interaction force is of the same magnitude as the
pinning force, but of opposite sign. Hence, the total force is zero and
the first kink is released. The second becomes pinned for a while until
it is released by the third and so forth. For non-adiabatically moving kinks,
the successor does not even become pinned by the predecessor, it only
lowers its velocity before the predecessor escapes due to the kink-kink
interaction and the successor becomes pinned at the boundary.

So far we have determined the frustration strength
$ \chi_{c}(0) $ above which
discommensurations enter the system in the absence of a driving force, $ f=0$.
In the presence of a driving force $ f$, we can roughly distinguish between
the regime above equilibrium threshold, $|\chi| > \chi_{c}(0)$, and below
equilibrium threshold $|\chi| < \chi_{c}(0)$.
Above equilibrium threshold, $|\chi| > \chi_{c}(0)$, depending on the
sign of $ \chi $, kinks or anti-kinks are present in the system, since
the entry barrier for discommensurations has vanished.
Neglecting the effects of the Peierls-Nabarro barriers, the threshold force
has basically vanished,
\begin{equation}
f_{c} \approx  0.
\end{equation}

For $|\chi| < \chi_{c}(0)$ there are no kinks
present in the system at equilibrium due to the finite entry barrier.
However, at a sufficiently large driving force, the entry barrier vanishes,
too. Then discommensurations enter the system at one boundary,
freely flow through it and exit at the other boundary.
In presence of a force the formation of kinks is similar to
the kink-penetration at equilibrium we discussed in Section \ref{CIT}.
Of course, the extremal solutions which determine the energy barrier for kink
formation are different. The stable and unstable solutions of the sine-Gordon
model in the presence of a driving force $f$ read \cite{Buet}
\begin{equation}
 u_{s,n} = a_{0} n + q^{-1} \arcsin(f/\mu)
\end{equation}
and
\begin{equation}
 u_{u,n} = \frac{(2n+1) a_{0} }{2} -  q^{-1} \arcsin(f/\mu),
\end{equation}
respectively.

For frustrations close to the equilibrium threshold frustration,
$ \chi_{c}(0) - |\chi| \ll \chi_{c}(0)$,
the depinning threshold force is small, $ f_{c} \ll \mu $.
Let us study the depinning of an anti-kink at the left boundary $ x=0$
for negative frustrations, $\chi < 0$, in the
presence of a small force, $ 0 < f \ll \mu$.
In the low-force limit, we can neglect the deformation of the anti-kink
due to the force. The energy of the driven anti-kink relative to the uniform
solution as a function of the anti-kink center $ x_{c} $ is then
\begin{eqnarray}
\Delta V(x_{c}) &\approx& \frac{{\tilde \kappa}  \chi }{a_{0}}
  [u_{a,n}(0;x_{c})- u_{a,n}(L;x_{c})] \\
&+& \frac{1}{a_{0}}   \int
dx \left\{ {\tilde \kappa}
\left[ \frac{\partial u_{a,n}(x;x_{c})}{\partial x} \right]^{2}
- f  u_{a,n}(x;x_{c}) \right\} \nonumber
\end{eqnarray}
If we neglect the presence of the other boundary at $ x=L$,
the derivative of the potential with respect to $ x_{c} $ is
 \begin{eqnarray} \nonumber
\frac{d}{d x_{c}} \Delta V(x_{c}) &\approx&
- \frac{{\tilde \kappa}}{a_{0}}
\frac{\partial  u_{a,n}(0;x_{c})}{ \partial x_{c}}
\left[\chi  + \frac{\partial u_{a,n}(0;x_{c})}{\partial x_{c}}  \right] \\
&+& \frac{f}{a_{0}} ~ u_{a,n}(0;x_{c})
\end{eqnarray}
The anti-kink depins at the left boundary when the maximum slope of the
potential at  $ x_{c}=0$ vanishes.
This occurs at the threshold force
 \begin{equation}
f_{c} = \frac{4 \mu}{\pi}\left[1- \frac{|\chi|}{\chi_{c}(0)} \right],
\end{equation}
which is easy to show realizing that $  u_{a,n}(0;0) = a_{0}/2 $,
$ \partial u_{a,n}(0;0)/ \partial x_{c} = \chi_{c}(0) $, and
$ \mu = \pi {\tilde \kappa} \chi_{c}^{2}(0)/(2 a_{0}) $.
The same result is found for kink depinning at the other boundary, $x=L$,
for positive frustration, $\chi > 0$.

\begin{figure}[t]
\begin{center}
  \epsfig{file=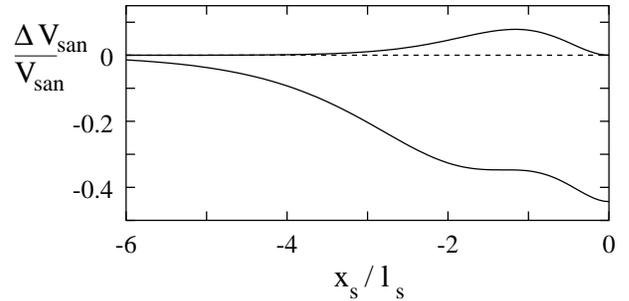,width=8cm}
\end{center}
\vspace*{-0.2cm}
\caption[]{\label{sanpot} Potential energy of a small amplitude nucleus
 $ \Delta V_{san} $ in units of
$ V_{san} = 4 {\tilde \kappa} a_{s}^{2}/15 a_{0} l_{s} $ as a function of
the position $ x_{s} $ of the maximum amplitude. Shown are values
close to the system boundary at $ x=0 $ for
frustration parameters $ \chi=-2 \sqrt{3} \chi_{c}(f)/5 $
(upper curve) and
$ \chi = \chi_{c}(f) $ (lower curve), where the barrier vanishes. }
\end{figure}

At low frustration, $ |\chi| \ll \chi_{c}(0)$, the depinning threshold
$f_{c}$ is close to $ \mu $. One thus has to consider the large force regime,
$ \mu - f \ll \mu, $ where the lowest energy saddle-point solution
of the sine-Gordon model $ u_{san,n}(x;x_{s}) =  u_{s,n} + \Delta u(x;x_{s}) $
has a small amplitude and hence is called small amplitude nucleus (SAN).
It can be calculated by approximating the tilted cosine potential by a cubic
parabola
\begin{equation} \label{san}
 \Delta u(x;x_{s}) = a_{s}
\cosh^{-2} \left[ \frac{x-x_{s}}{2 l_{s}} \right],
\end{equation}
with center $ x_{s}$, amplitude $ a_{s}= 3 q^{-1} [2(1-f/\mu)]^{1/2}$,
and width $ l_{s} = l_{d} [2(1-f/\mu)]^{-1/4} $.

\begin{figure}[t]
  \begin{center}
    \epsfig{file=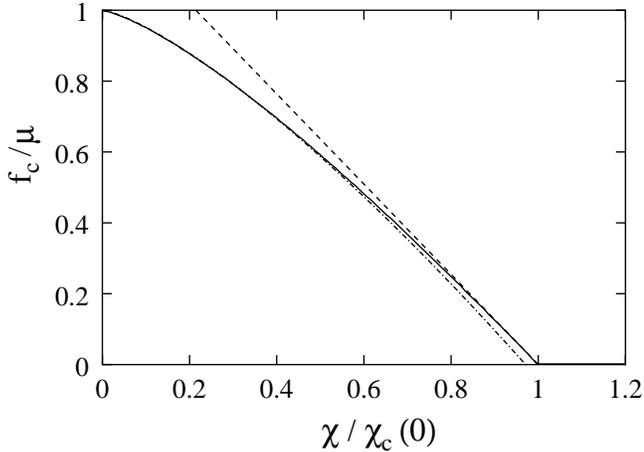,width=8.5cm}
  \end{center}
  \vspace*{-0.2cm}
  \caption[]{  \label{frustrat}
Critical force $ f_{c} $ as a function of the mismatch parameter $ \chi$.
Shown are the numerical integration results (solid line),
high force approximation $ f_{c}-\mu \ll \mu$ (dashed-dotted line) and the
low force approximation $ f_{c} \ll \mu$ (dashed line). }
\end{figure}
%%%%%%%%%%%%%%%%%%%%%%%%%%%%%%%%%%%%%%%%%%%%%%%%%%%%%%%%%%%%%%%%%%%%%%%%%%%%

Let us now consider the penetration of an anti-kink for $ \chi < 0 $ at
the boundary $ x=0 $.
The energy difference $ \Delta V_{san} = V[u_{san,n}] - V[u_{s,n} ] $ as
a function of $ x_{s} \le 0 $ is
\begin{eqnarray} \nonumber
&& \Delta V_{san} (x_{s}) =
\frac{4 {\tilde \kappa} a_{s}^{2}}{15 a_{0} l_{s}}
   \left\{ 1 + \frac{5}{2 \sqrt{3}}
             \frac{\chi}{\chi_{c}(f)}
                 \frac{\Delta u(0,x_{s})}{a_{s}} \right. \\ && \left.
 - \left[ \frac{3 \Delta u(0,x_{s})}{2 a_{s} } + 1\right]
           \left[1- \frac{ \Delta u(0,x_{s})}{a_{s} }\right]^{3/2}
   \right\},
\end{eqnarray}
where
\begin{equation}
  \chi_{c}(f) = \frac{\pi \chi_{ci}  }{2 \sqrt{3}}
\left[ 2 \left(1-\frac{f}{\mu} \right)
\right]^{3/4}.
\end{equation}
As shown in Fig.\ \ref{sanpot}, the SAN potential $  \Delta V_{san}(x_{s}) $
has a barrier (upper curve), which vanishes at $ \chi = -\chi_{c}(f)$ (lower curve).
This can be seen by analyzing the zeros of the derivative
\begin{eqnarray} \nonumber &&
\frac{d}{d x_{s}} \Delta V_{san} (x_{s}) =
\frac{ {\tilde \kappa} a_{s}}{ a_{0} l_{s}}
\frac{\partial \Delta u(0,x_{s})}{\partial x_{s}}
   \left\{ \frac{2}{3\sqrt{3}} \frac{\chi}{\chi_{c}(f)} \right. \\
&& \left. +
\frac{ \Delta u(0,x_{s})}{a_{s} }
         \left[ 1-
\frac{\Delta u(0,x_{s})}{a_{s} } \right]^{1/2}
   \right\}.
\end{eqnarray}
One zero is given by $ x_{s}=0 $, where the partial derivative
$\partial \Delta u(0,x_{s})/\partial x_{s} $
vanishes. The others can in principle be found by studying the term in the
curly brackets which becomes zero if
\begin{equation} \label{curly}
  \left[\frac{ \Delta u}{a_{s}} \right]^2
         \left[1- \frac{\Delta u}{a_{s} } \right]
         = \frac{4}{27} \left[ \frac{\chi}{\chi_{c} (f)} \right]^2.
\end{equation}
However, to find the threshold condition, we do not need to calculate
$ x_{s} $ explicitely, it is sufficient to determine $ \Delta u(0;x_{s})$.
Equation (\ref{curly}) has at most three roots, depending on the value of
$ \chi^2>0$. One of the roots is negative, which is
no solution, since $ 0<\Delta u \le a_{s} $ [see Eq. (\ref{san})].
For $ -\chi_{c}(f) < \chi < 0$
there are two positive roots, which indicate the existence of two extrema
of $ \Delta V_{san}$, a minimum and a maximum. Calculating the
extrema of Eq.\ (\ref{curly}) with respect to $ \Delta u$, one sees that
at $ \chi= -\chi_{c}(f) $ the positive roots of Eq.\ (\ref{curly})
become degenerate,
which means that the minimum and the maximum merge into a saddle point
of $ \Delta V_{san}$. Hence, the entry barrier of the SAN vanishes.
At this value,  $\Delta u=2a_{s}/3$.
Thus, two thirds of a SAN are localized at
the left boundary, $ x=0$, but are unstable against small perturbations.
Increasing either  $ \chi $ or $ f $ depins the SAN which then evolves into
a full anti-kink. Finally, from $\Delta u(0;x_{s})=2a_{s}/3$ we find
$x_{s}= l_{s} \ln(2 - \sqrt{3})$.

To summarize, for low frustrations $ | \chi| \ll \chi_{c}(0) $
and for frustrations close to the equilibrium threshold
$ \chi_{c}(0)-|\chi| \ll \chi_{c}(0) $
the threshold force $ f_{c} $ is given by
\begin{equation} \label{frustthresh}
 f_{c} \approx \left\{
                 \begin{array}{ll}
  \mu \left\{1-\frac{3^{2/3}}{2}
     \left[ \frac{ |\chi|}{\chi_{c}(0)} \right]^{4/3}  \right\},
                          &    \quad |\chi| \ll \chi_{c}(0)\\
                          & \\
  \frac{4}{\pi} \mu  \left[1- \frac{|\chi|}{\chi_{c}(0)} \right],
                          &   \quad \chi_{c}(0)-|\chi| \ll
     \chi_{c}(0)\\
&\\
   0,                       &   \quad |\chi| \gsim \chi_{c}(0).
                  \end{array}
         \right.
\end{equation}

For completeness, we calculated the threshold force for arbitrary frustration
by numerical integration.
To calculate the static and dynamic solutions of Eq.\ (\ref{CFKM}),
we use a standard numerical integration procedure.
Starting with a flat initial
configuration, $ u_{m} = 0$, we iterate
\begin{equation} \label{numeric}
u_{m}(t+\delta t) =  u_{m}(t) + \delta t ~ v_{m}(t),
\end{equation}
with
\begin{eqnarray} \nonumber
v_{m}(t) &=&  f + \sin \left[2 \pi u_{m}(t) \right] +
2 \pi g  [u_{m+1}(t) + u_{m-1}(t) \\
&-& 2u_{m}(t)
+ \delta_{m,M}-\delta_{m,1}]
\end{eqnarray}
where length is measured in units of $ a_{0} $,
time in units of $ t_{0}= a_{0}\eta/\mu$,
and force in units of $ \mu $. The length of the system is
$ M = L/a_{0} $. Recalling that the vortex-vortex
interaction energy falls off exponentially fast for distances between vortices
larger than $ \lambda $, we take only the $ N = [5\lambda/a_{0} ]$ next
neighboring vortices into account in the sum over channel vortices.
The channel has a length of $ L $.
At its ends, we apply boundary conditions taking into account the
frustration, $ \chi $.
For a given force, Eq.\ (\ref{numeric}) is iterated until a fairly steady
state is reached, $ [v_m(t+\delta t)-v_m(t)]/\delta t < 10^{-4} $.
Channel vortices are defined to be static, if $ v_m < 10^{-8}.$
The calculated $ u_{m}(t) $ and $ v_{m}(t) $ are recorded for several
forces. In addition we can record
the particle trajectories $ x_{m}(t)$ to visualize the dynamical behavior of
the channel vortices close to the depinning transition.
The numerical results and the analytical limits for $ f_{c}(\chi) $ are
shown in Fig.\ \ref{frustrat}.
\begin{figure}[t]
\begin{center}
  \epsfig{file=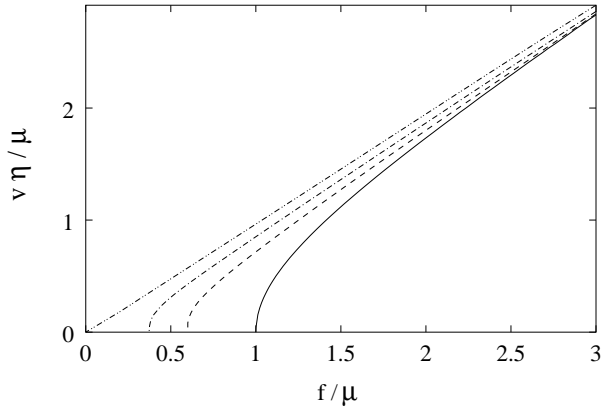,width=8cm}
\end{center}
\vspace*{-0.2cm}
\caption[]{\label{vfcurveord}
           Typical $f-v$ curves computed for  systems with
           $L=1000 a_{0}$ and $\lambda=a_{0}$. The result for the
           unfrustrated commensurate case, $ \chi=0 $ (solid line),
           is identical to the
           curve of a single particle in a sinusoidal potential,
           $v \sim (f-\mu)^{\nu} $ with $ \nu = 1/2 $.
           Frustration lowers the threshold force
           $f_{c}$, but does not alter the exponent $ \nu $
           in the commensurate regime. However, in the incommensurate
           state, $ |\chi| > \chi_{c}(0)$, the $f-v$ curve becomes linear,
           $ v \sim f$.
           Shown are the $f-v$ characteristics for frustrated systems with
           $ \chi=\chi_{c}(0)$   (dash-dot-dotted), $0.7~ \chi_{c}(0) $
           (dash-dotted), and $0.5~ \chi_{c}(0)$ (dashed).
          }
\end{figure}

The numerical integration also allows to determine the velocity averaged
over time and space,
\begin{equation}
v=\frac{a_{0}}{L}\sum_{m=1}^{L}\int_{0}^{T}\frac{dt}{T}~ {\dot u}_{m}(t),
\end{equation}
as a function of the force $f$ as
shown in Fig.\ \ref{vfcurveord}.

Above $ \chi_{c}(0) $ where the entry barrier has vanished
the topological defects move freely through the sample. The
linear force-velocity characteristics resembles to the one of
a single free particle
with dissipative dynamics, $ v = f/\eta$. The velocity of the entire chain
is determined by the velocity of the defects that enter the system
at the boundary. At $ |\chi|= \chi_{c} $, where the entry barrier for
defects has vanished, $ \partial_{x}^{2} u(x=0,L) = 0 $. This means that close
to the maximum of the sinusoidal potential, where the chain bead
spends most of the time, it effectively behaves like a single particle which
is driven by a force $f$. For $ |\chi|> \chi_{c} $ a similar argument
can be given. At $ f=0$ the effective force at the system boundary pushes
defects into the channel
until their density is so high, that their repulsion prevents new defects
to flow in. Effectively, the chain bead at the boundary reaches an unstable
equilibrium. Driving the system now with a non-zero force $f>0$
results in the same  motion for the bead at the boundary as for
$ |\chi|= \chi_{c} $.

For $ \chi < \chi_{c}(0)$ the
entry barrier for the defects becomes relevant and the force-velocity
characteristics shows the behavior typical for quasi-particles with a
vanishing saddle-point barrier with $ v \sim [f- f_{c}(\chi)]^{1/2}$
for $ f \to f_{c}$. Finally,
in the absence of frustration the particles depin instantly and
 $ v = (f_{c}/\eta) (2f/f_{c}-2)^{1/2} $
like a single particle in a sinusoidal potential.

\subsection{Preliminary comparison with experimental data}

Let us now compare these theoretical findings with experimental and
numerical data.\cite{Bess}
Clearly, the generalized Frenkel-Kontorova model predicts a nearly
vanishing depinning force $f_{c} \approx 0 $ in the defective state.
In the commensurate state, the depinning threshold $ f_{c} $ is finite, but
decreases monotonously with increasing frustration and vanishes at the
commensurate-incommensurate transition,
where the critical mismatch is reached, $ \chi = \chi_{c}(0) $, see
Fig.\ \ref{frustrat}.
As a typical value for $\lambda/a_0$ in the low field limit we consider
$ \lambda = 3 a_{0}$. From Eq.\ (\ref{g}) we
find $ g \sim 28 $ and hence Eq.\ (\ref{chi_c}) yields
$ \chi_{c}(0) \sim 0.06 $. Since $ \chi = (b_{0}/w) -1$, one expects a sharp
spike in the threshold force around the magnetic induction with a vortex
lattice plane distance $ b_{0} = w $.

If we associate the maxima of the critical force to the spikes corresponding
to a commensurate state and the minima to the incommensurate one, we
would nonetheless observe that some essential ingredient must be missing
in our model. Indeed, in Fig.\ 3 we see that instead of sharp spikes, the
measurements rather show a continuous modulation of the critical current
as a function of the magnetic field, or equivalently, as a function of
the frustration inside the channel. In addition, the data do not exhibit
the vanishing pinning barriers which are expected for incommensurate
(defective) structures with perfectly ordered channel edges. Both phenomena
cannot be explained solely in terms of thermal or quantum fluctuations:
They would lead to an effective reduction of the pinning barrier in both
cases. This motivates us to study the influence of quenched disorder
in the environmental vortex lattice. Actually, recent experiments have
unvealed the essential role of disorder. It turns out that the picture of
maximum
critical force corresponding to an integer number of rows inside the channel
is only valid for zero and weak disorder. If the disorder is strong, exactly
the opposite will be realized, namely, the integer number of rows will
correspond to the minima of $f_c$ (Ref.\ [\cite{rutlong}]).

It is generally understood, that quenched disorder leads to pinning of vortices
\cite{Blat}
or topological defects of the vortex lattice, which would explain the
increased critical force in the defective state. It remains to understand
how disorder leads to a reduction of pinning barriers in the commensurate
regime.

\subsection{Channel with a distorted vortex lattice in the environment}

In a realistic sample the static nature of the channel environment is caused by
some sort of pinning which may distort the vortex lattice.
In the following we consider a device, where the vortices outside the channel
are pinned by quenched disorder.
The pinning in the channel environment is strong enough to guarantee
that it remains pinned at all considered current densities
whereas inside the channel pinning by quenched disorder is orders of
magnitudes lower
and can be entirely neglected. In a typical experiment the magnetic
inductions are so large that the interaction between the
vortices is much stronger than the pinning. Further, we assume that the
vortex lattice in the channel environment is free of dislocations. Then it
is natural to treat the vortex lattice in its elastic limit.
In a weakly pinned vortex lattice, distortions of the order of the
coherence length $ \xi $ due to the disorder
occur on average on a length $ l_{c} $  in the direction of the magnetic
field and on a length $ R_{c} $ transverse to it.
Here, we consider  two-dimensional  collective pinning where
$ l_{c} \gg l $ such that the approximation of
straight vortex lines remains valid.

With quenched disorder in the environment of the channel,
the interaction energy of a single vortex with the vortex row becomes
\begin{eqnarray}
E_{n}({\bf r})&=&\sum_{m} U({\bf r}- {\bf R}_{m,n} - {\bf d}_{m,n} ),
\end{eqnarray}
where
$ {\bf d}_{m,n}$ are the displacements from the ordered positions
$ {\bf R}_{m,n}$.
It is convenient to rewrite the row potential in terms of the vortex
density
\begin{eqnarray}
E_{n}({\bf r})&=&\int d^{2} r'  U({\bf r}-{\bf r'}) \rho_{n}({\bf r}'),
\end{eqnarray}
where
\begin{eqnarray}
\rho_{n}({\bf r'})&=&\sum_{m} \delta({\bf r'}- {\bf R}_{m,n} - {\bf d}_{m,n} ).
\end{eqnarray}
We introduce a continuous displacement field
\begin{eqnarray}
{\bf d}({\bf r})
&=& \frac{a_{0}^{2}}{4 \pi^{2}} \int_{BZ} d^{2} k~ e^{i {\bf k}{\bf r}}
\sum_{m,n} e^{-i {\bf k}{\bf R}_{m,n}} {\bf d}_{m,n},
\end{eqnarray}
where $BZ$ indicates that the integration is restricted to the first
Brillouin zone. Note that $ {\bf d}({\bf R}_{m,n}) = {\bf d}_{m,n} $.
In order to derive a simplified one-dimensional model as in the ordered
case, we make use of a few approximations.
Since we consider the vortex lattice to be in the elastic limit, then
\begin{equation}  \label{elacon}
 | \nabla \cdot {\bf d}| \sim \xi / R_{c} \ll 1.
\end{equation}
Moreover, since the vortex potential  falls off exponentially fast
for $|{\bf r} - {\bf r'}| > \lambda, $
the channel environment is mainly probed within
$|{\bf r} - {\bf r'}| < \lambda$, where one can estimate
$ |{\bf d}({\bf r'})- {\bf d}({\bf r})| \lsim \lambda \xi / R_{c} $
using Eq.\ (\ref{elacon}).
For  $ \lambda \xi / R_{c} \ll a_{0}/2$ it is then reasonable to expand
the displacements $ {\bf d}({\bf r'}) $ in the integral,
\begin{equation}
{\bf d}({\bf r'}) = {\bar {\bf d}}({\bf r})
+ {\cal O}\left( \frac{\lambda \xi}{R_{c}} \right).
\end{equation}
Here, we have introduced the coarse grained displacement
field
\begin{equation}
  {\bar {\bf d}}({\bf r}) =  (2 \pi U_{0})^{-1} \lambda^{-2}
\int d^{2} r' U({\bf r}-{\bf r'})
{\bf d}({\bf r'}),
\end{equation}
which is smooth on the scale $ \lambda $.
Up to terms of the order
$ {\cal O}( \lambda \xi/R_{c} )$, we obtain\cite{rutlong}
\begin{equation}
E_{n}({\bf r}) = E_{n}^{o} [{\bf r} - {\bar {\bf d}}({\bf r})]
\end{equation}
for the collectively pinned vortex row potential.

To calculate the effective channel potential for a channel of
width $ w \sim b_{0}, $
we perform the summation over pinned vortex rows $E_{n}$ as
in the ordered case.
Further, since the influence of point-like disorder at the edge is
assumed to be much
weaker than the re-ordering due to the edge currents, we may well take
$ {\bar d}_{y}({\bf r}) =0$. Then,
the interaction of a single vortex in the
channel with the disordered environment reads
\begin{equation}
E_{dc}({\bf r}) =   \sum_{\nu} A_{\nu}(y)
 \cos \left\{q_{\nu}[x-{\bar d}_{x}({\bf r})] \right\},
\end{equation}
For the partial derivatives of the channel potential one finds
\begin{eqnarray} \nonumber
\partial_{x} E_{dc}({\bf r})& = & - \sum_{\nu} q_{\nu} A_{\nu}(y)
 \sin \left\{q_{\nu}[x-{\bar d}_{x}({\bf r})] \right\}
\\
\partial_{y} E_{dc}({\bf r})& = &
 \sum_{\nu} A_{\nu}'(y)
 \cos \left\{q_{\nu}[x-{\bar d}_{x}({\bf r})] \right\} \nonumber
\end{eqnarray}
plus terms of order ${\cal O}\left( \lambda \xi / R_{c} \right)$.
As in the perfectly ordered case, we now consider the equation of motion.
Substituting $ E_{oc} $ by $ E_{dc} $ in Eq.\ (\ref{equofmot}) and following
similar arguments we derive an equation for the longitudinal  motion.
Introducing $ {\tilde \varphi}(x) = q {\bar d}_{x}(x,0) $ we obtain
a generalized phase-disordered Frenkel-Kontorova model,
\begin{eqnarray} \nonumber
 V &=& \sum_{m} \left\{
 \frac{\mu}{q} \left\{1- \cos \left[ q x_{m}-{\tilde \varphi}(x_{m})
                              \right] \right\} - f x_{m} \right.  \\  \label{DFKM}
&+& \left. \sum_{n} \frac{\kappa_{n}}{2} (x_{m+n} - x_{m} -na)^{2} \right\}.
\end{eqnarray}
The corresponding energy functional in the continuum limit is then
\begin{eqnarray} \nonumber
 V[u] &=& a_{0}^{-1} \int_{0}^{L} dx \left\{
\frac{{\tilde \kappa}}{2}
\left(\frac{\partial u}{\partial x}-\chi \right)^{2} +
 \frac{\mu}{q} \left\{1- \right. \right. \\  && \left. \left.
\cos \left[ q u - {\tilde \varphi}(x+u) \right]
\right\}
- f u
\right\} \label{VDFKM}
\end{eqnarray}
and the resulting equation of motion for the displacement fields $ u(x,t) $
reads
\begin{eqnarray} \nonumber
{\eta} {\dot u} &=&
f - \mu \sin \left[ q u-{\tilde \varphi}(x+u) \right]
+  {\tilde \kappa} \frac{\partial^2 u}{\partial x^2 } \\
&+&  {\tilde \kappa} \chi[\delta(x-L)-\delta(x)],\label{CDFKM}
\end{eqnarray}
where  $ \partial_{x} {\bar d}_{x}$-terms that are of order
$  {\cal O}( \lambda \xi / R_{c}) $ have been neglected.

\subsection{Depinning in a channel with distorted environment}
\label{distort}
Up to now, we have not specified the disorder displacement field
$ {\bar d}_{x}$. To gain some basic understanding, we now consider the
effect of a local distortion on the depinning properties.
The field we choose is somewhat academic, but is convenient to understand the
effect of a lattice distortion at the system boundary and in the bulk.
The perturbation occurs
around $ x_{\varphi}$,
\begin{equation} \label{distortion}
 {\bar d}_{x}(x,0) =  W (x-x_{\varphi})
 \left[ \theta(x-x_{\varphi})-\frac{1}{2} \right].
\end{equation}
Here, $ \theta(x) $ is the Heaviside function and $ W > 0$ the distortion
parameter.
It is convenient to introduce transformed displacement fields
$ {\tilde u} = u - q^{-1} {\tilde \varphi}$.
Neglecting terms of the order
$(\partial_{x} {\tilde u})^{2} {\tilde \varphi}'$and
$({\tilde \varphi}')^{2}\partial_{x} {\tilde u} $,  the relevant
$ {\tilde u}$-dependent part of the energy functional reads
\begin{equation}
{\tilde V}[{\tilde u}] =
V_{SG}[{\tilde u}]+V_{\chi}[{\tilde u}]+V_{\varphi}[{\tilde u}],
\end{equation}
where
\begin{equation}
V_{\varphi}[{\tilde u}] = \frac{\tilde \kappa}{a_{0}}
\int dx \frac{\partial {\tilde u}}{\partial x}
\frac{\partial {\bar d}_{x}}{\partial x}.
\end{equation}
In the following, we examine the effect of lattice distortions at the
boundaries, $ x_{\varphi} = 0,L$, and in the bulk, $ x_{\varphi}=L/2 $.
To gain a basic understanding of the depinning process, we restrict
the analysis to the large system limit, $ L/2 \gg l_{s} $, where the
system is so large that the depinning configurations at the weak spots
$ x_{\varphi} = 0,L,$ and $ x_{\varphi}=L/2 $ do not interact with each
other.

For $ x_{\varphi} = 0$, the lattice is distorted
homogeneously in the entire sample
and the contribution of the distortion to the energy functional
yields
\begin{equation} \label{distortionenergy}
V_{\varphi}[{\tilde u}] = \frac{{\tilde \kappa} W}{2 a_{0}}
[{\tilde u}(L)-{\tilde u}(0)].
\end{equation}
Physically, the distortion results in an additional frustration of the
system, as can be immediately understood by comparing
Eq.\ (\ref{distortionenergy}) with Eq.\ (\ref{frustrationenergy}).
This means that for the threshold force, one can use the results
that were found
in the absence of the distortion [see Eq.\ (\ref{frustthresh})],
but the frustration has
to be replaced by an effective frustration
\begin{equation}
  \chi  \to \chi-\frac{W}{2}.
\end{equation}
The result is thus a simple shift of the $\chi-f_{c}$-curve.
Similarly, for $ x_{\varphi} = L $, the $ f-\chi$-curve is shifted,
\begin{equation}
  \chi  \to \chi+\frac{W}{2}.
\end{equation}

For $ x_{\varphi} = L/2$, in addition to boundary depinning, bulk depinning
at $  x_{\varphi} $ can occur. In large systems,
$ L/2 \gg l_{s} $ we can treat the effect of the distortions
at the boundaries and in the bulk
separately. At the boundaries, the threshold solution
is then approximately given by the solution in absence of the defect, but
with an increased  effective frustration $\chi+W/2$ in the left half and a
lowered  effective frustration $\chi-W/2$ in the right half of the system.
The threshold force for boundary depinning is again given by
Eq.\ (\ref{frustthresh}) with the modulus of the frustration replaced by
\begin{equation}
  |\chi| \to \left| |\chi|-\frac{W}{2} \right|.
\end{equation}
To understand bulk depinning, we first restrict the considerations to
$ \chi=0$ and then discuss the behavior in the presence of frustration.

For $ L/2 \gg l_{s} $ we can neglect the influence of the boundaries and treat
them as if they were shifted to $ \pm \infty $. Then,
 the extremal threshold solution
$ {\tilde u}_{\varphi}(x;x_{\varphi}) $ can be constructed by joining
two  extremal solutions of $ V_{SG}$. The matching
condition is found from
\begin{equation}
\frac{\partial u(x-\varepsilon)}{\partial x} =
\frac{\partial u(x+\varepsilon)}{\partial x}
 \end{equation}
such that the matching condition for the transformed field
$ {\tilde u}_{\varphi}(x;x_{\varphi}) $ reads
\begin{equation}
\frac{\partial  {\tilde
    u}_{\varphi}(x_{\varphi}-\varepsilon;x_{\varphi})}{\partial x} -
\frac{\partial  {\tilde
    u}_{\varphi}(x_{\varphi}+\varepsilon;x_{\varphi})}{\partial x} = W
\end{equation}
The mirror symmetry
requires $ \partial_{x} u(x_{\varphi})=0 $, hence
$ \partial_{x} {\tilde u}(x_{\varphi} \pm \varepsilon)=
\mp W/2 $.
At the threshold force $ f_{c}$ stable solutions cease to exist. In fact,
it can be shown that at $ f_{c} $ the stable solution merges with an unstable
one. This occurs, when the maximum of the tongue  developing at
$ x_{\varphi} $ reaches the maximum of the sinusoidal potential.
>From $ \delta V/\delta u=0 $  follows that at the potential maximum
the extremal solution has to fulfill $ \partial_{x}^{2} u(x_{\varphi})=0$,
which holds if
$ \partial^{2}_{x}{\tilde u}(x_{\varphi} \pm \varepsilon) \to 0$
for $ \varepsilon \to 0 $.

For weak distortions, $ |W/2| \ll 1/(\pi \sqrt{g})$, the threshold force
for bulk depinning is close to $ \mu $ and the threshold solution
$ {\tilde u}_{\varphi} = {\tilde u}_{s} + \Delta {\tilde u}_{\varphi}$
can be found by merging two  SAN solutions at $ x_{\varphi} \pm x_{a}$,
where $ x_{a} = l_{s} \ln(2+\sqrt{3})$,
\begin{equation}
 \Delta {\tilde u}_{\varphi}(x;x_{\varphi}) =
\left\{ \begin{array}{ll}
\Delta u (x;x_{\varphi}+x_{a}) & x <x_{\varphi}\\
\Delta u (x;x_{\varphi}-x_{a}) & x \ge x_{\varphi}.
\end{array}
\right.
\end{equation}
The maximum value of the tongue  developing at $ x_{\varphi} $
is given by
$ \Delta {\tilde u}_{\varphi}(x_{\varphi};x_{\varphi}) =2 a_{s}/3 $.
This implies
\begin{equation}
\frac{\partial \Delta {\tilde
    u}_{\varphi}(x_{\varphi}-\varepsilon;x_{\varphi})}{\partial x} =
    \frac{2 a_{s}}{3 \sqrt{3} l_{s}} =\frac{W}{2},
\end{equation}
from which one obtains the bulk depinning threshold force
in the presence of weak distortions,
\begin{equation}
{\hat f}_{c} =
\mu \left[1 - \frac{1}{2} \left(
\frac{ \pi \sqrt{3g}}{2}~ W \right)^{4/3}
\right].
\end{equation}
This formula becomes invalid for
$  2/(\pi \sqrt{g}) -|W|\ll  2/(\pi \sqrt{g})$.
For strong distortions, $ |W| \lsim 2/(\pi \sqrt{g})$, the threshold
configuration can be constructed by merging a kink
and an anti-kink
 \begin{equation}
{\tilde u}_{\varphi}(x;x_{\varphi}) =
\left\{ \begin{array}{ll}
u_{k}(x;x_{\varphi}) & x <x_{\varphi}\\
u_{a}(x;x_{\varphi}) & x \ge x_{\varphi}
\end{array}
\right.
\end{equation}
from which one obtains the bulk depinning threshold force
in the presence of strong distortions,
\begin{equation}
{\hat f}_{c} =  \frac{4 \mu}{\pi} \left(1-  \frac{ \pi \sqrt{g}}{2}~ W \right).
\end{equation}

At $ W = 2/(\pi \sqrt{g})$ stable saddle-point solutions cease to exist
for all $ f $ and
disorder induced mobile kink-anti-kink pairs are spontaneously formed
even at equilibrium.

After having gained some understanding how pinning occurs at a weak spot in
the bulk for $\chi=0$,
let us now consider the frustrated case, $ |\chi| >0 $,
where bulk depinning competes with boundary depinning.
Comparing the bulk depinning threshold $ {\hat f}_{c}(W) $
with the  boundary threshold in presence of the defect, $ f_{c}(|\chi|-W/2)$,
we find that for $ |\chi| < W $ the system depins in the bulk and
for $ |\chi| > W $ at one of the boundaries.
Note that to lowest order, we can apply these results to distortions of this
kind that are not necessarily centered at $ x_{\varphi} = L/2$, as long as
$ l_{s} \ll x_{\varphi} \ll L - l_{s} $ holds.

\subsection{Depinning in a channel with randomly displaced edge vortices}
Opposed to the rather well behaved distortions of the previous paragraph,
we now consider  the effect of randomly displaced vortices in the channel
edge. We mimic the disorder by uncorrelated relative displacements
$ [{\bar d}_{x}({\bf R}_{m+1,n}) - {\bar d}_{x}({\bf R}_{m,n})]/a_{0} $.
The latter are independent identically box-distributed random numbers
within the interval $ [-W/2,W/2]$,
\begin{equation}
P_{W}({\bar d}_{x})= \frac{1}{W} \left[\theta({\bar d}_{x}+W/2)
                                       -\theta({\bar d}_{x}-W/2) \right].
\end{equation}

The width of the box distribution $W$ parameterizes the disorder strength.
Then, ${\tilde \mu}$ and $ \partial_{x} {\tilde \varphi}$ are random functions, which are smooth on the scale
$ \lambda $ and bounded. Note that although $\partial_{x} {\tilde \varphi}$
is bounded, ${\tilde \varphi}$ is unbounded. Thus, long range order is lost
along the channel direction. On length scales much larger than $ a_{0}$,
the displacement field ${\bar d}_{x}({\bf r})$ behaves like a random walk
in 1D and the phase-phase correlator scales linearly with the distance,
$\langle[{\tilde \varphi}(x)-{\tilde \varphi}(0)]^2\rangle \propto x$.

\begin{figure}
\begin{center}
\epsfig{file=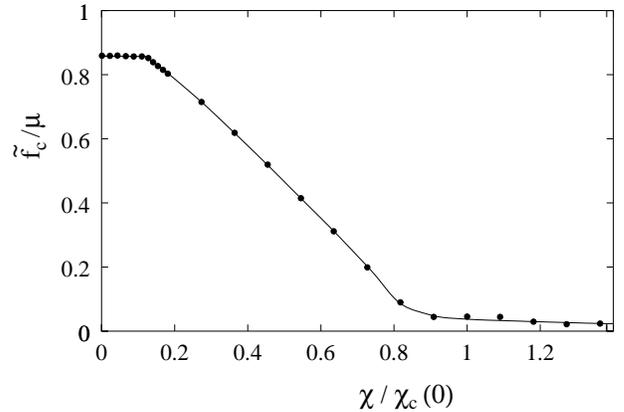, width=8.cm}
\end{center}
\caption{\label{num2}  Numerically
obtained minimum threshold force $ {\tilde f}_{c} $
as a function of frustration $ \chi $ for
$100$ channels with $ L=100 a_{0} $, $ \lambda=a_{0}$, and $W=0.1$.
}
\end{figure}

\begin{figure}[t]
  \begin{center}
    \epsfig{file=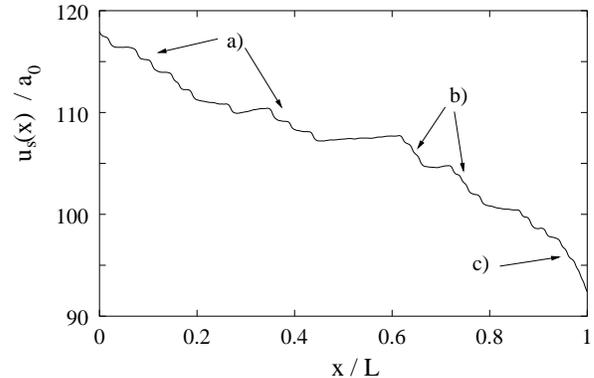,width=8cm}
  \end{center}
  \vspace*{-0.2cm}
  \caption[]
    {\label{istat}
     Static displacements $u_{s}(x)$
     in a system of length $L=1000 a_{0}$ for $ \chi=-\chi_{c}(0)$
     in the pinned regime, $ f < {\tilde f}_{c}$.
     (a) Pinned single interstitials (kinks). (b) Multiple pinned kinks
     at strong pinning sites. (c) Kink accumulation at the
     end of the system, $ x \lsim L $, where in addition to disorder
     the exit barrier is present.
   }
\end{figure}

The unfrustrated case has been discussed in Ref.\ [\cite{rutlong}].
Here we generalize the results for
$ |\chi|>0$.
In weakly frustrated systems, $ |\chi| <  W/2 $,
nucleation still occurs at a weak spot in the bulk. Hence, the
depinning threshold $ {\tilde f}_{c}$ should be independent of
the frustration $ \chi $. This can
be indeed observed, see Fig. \ref{num2}:
the $\chi-{\tilde f}_{c}$-curve has a plateau around $ \chi \sim 0$, before
at larger values of  $ \chi $ depinning takes place via the formation of
defects at the boundary as in the ordered case.
The threshold for boundary depinning
is affected by lattice distortions at the sample edge as discussed in Section
\ref{distort} resulting in an overall shift of the
$\chi-{\tilde f}_{c}$-curve to lower values of $ \chi $.

For intermediate frustration, $ 0 \ll \chi \ll \chi_{c}(0) $, a defect that
entered the sample via the boundary cannot be pinned by disorder in the bulk.
It moves freely to the other boundary, where
it becomes pinned by the exit barrier until being released by the next
defect that enters the channel and then travels freely to the exit.

\begin{figure}[t]
  \begin{center}
    \epsfig{file=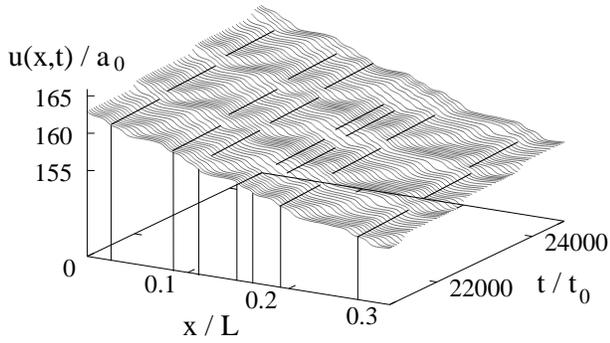,width=8cm}
  \end{center}
  \vspace*{-0.2cm}
  \caption[]
    {\label{idepin}
     Time evolution of the displacements $u(x,t)$
     in a system of length $L=1000 a_{0}$ for $ \chi=-\chi_{c}(0)$
     and $ f >{\tilde f}_{c}$.
     Strong pinning sites are indicated by bars parallel to the time axis.
     Here, interstitials (anti-kinks) travel to the right.
     At strong pinning sites they collide with pinned kinks. The formerly
     pinned kinks are released while the previously moving ones become
     pinned.
    }
\end{figure}

For frustrations above $ \chi_{c}(0) - W/2 $ bulk pinning becomes
possible since the boundary depinning threshold
force becomes lower than the disorder induced pinning forces in the bulk.
In Fig.\ \ref{num2} bulk pinning becomes relevant
around $ \chi \sim 0.8 \chi_{c}(0)$. Indeed one can observe pinning of
defects and bundles of defects in the numerical simulations in this regime.
In Fig.\ \ref{istat} a static state in a system of length $L=1000 a_{0}$
for $ \chi=-\chi_{c}(0)$ just below the depinning threshold
$ f  \lsim {\tilde f}_{c}$ is shown. One clearly sees single pinned kinks
and a few multiple-pinned kinks (bundles). At the exit of the system kinks
accumulate due to  both bulk pinning and due to pinning at the presence of
the exit barrier. When reaching the depinning threshold, a defect is formed
at the entry on the channel and travels until it collides with a defect that
is already pinned at a strong pinning site. While the latter becomes released,
the former gets pinned. This scenario repeats until a mobile
defect has reached the channel exit, see Fig.\ \ref{idepin}.

\begin{figure}[t]
\begin{center}
  \epsfig{file=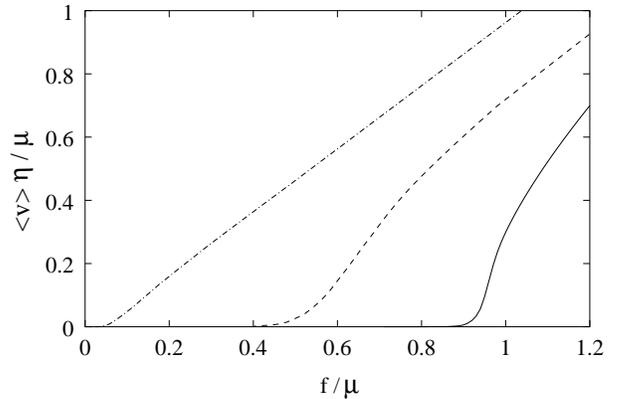,width=8cm}
\end{center}
\vspace*{-0.2cm}
\caption[]{\label{vfcurvedisord}
           Disorder averaged $f-v$ curves computed for  systems with
           $L=100 a_{0}$ and $\lambda=a_{0}$.
           Shown are the $f-v$ characteristics for systems with
           $ \chi=\chi_{c}(0)$   (dash-dotted),
           $\chi=0.5~ \chi_{c}(0) $
           (dashed), and $\chi=0$ (solid).
          }
\end{figure}

In contrast to the force-velocity curve of single systems, which
behave as $ v \sim (f-f_{c})^{\nu}$ with $ \nu=1/2$,
the functional form of the disorder
averaged force-velocity curves for finite  systems all show upward curvature
for $ f \gsim {\tilde f}_{c}$, which corresponds to  $ \nu>1$,
see Fig.\ \ref{vfcurvedisord}. A crossover
between exponents similar to the transition $ \nu=1/2 \to 1$ occurring in
 the ordered model at $ \chi_{c}(0)$ can not be observed in the disordered
case.

\begin{figure}
\begin{center}
\epsfig{file=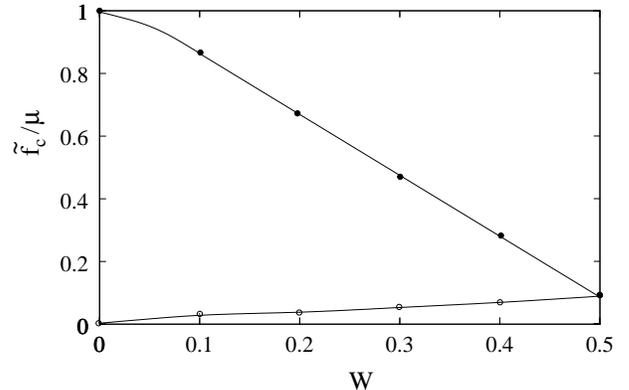, width=8.cm}
\end{center}
\caption{\label{num3}  Numerically
obtained minimum threshold force $ {\tilde f}_{c} $
as a function of disorder strength $ W $ for
$100$ channels with $ L=100 a_{0} $, and $ \lambda=a_{0}$.
Upper curve: $\chi=0$, lower curve $\chi=\chi_{c}(0)$.
}
\end{figure}

Until now we considered the weak disorder limit in agreement with the
assumptions made in order to develop the disordered model. If we increase the
disorder parameter $W$ beyond the initially assumed limits, we can gain some
insight into the depinning properties at large disorder.
In Fig.\ \ref{num3}  minimum threshold forces $ {\tilde f}_{c} $
as a function of disorder strength $ W $ for
$100$ channels with $ L=100 a_{0} $, and $ \lambda=a_{0}$
are shown for systems without frustration $\chi=0$ and for
frustrated systems with $\chi=\chi_{c}(0)$, where we expect the effect of the
boundaries to become irrelevant. At $ W=0.5$ the disorder is so strong,
that a distinction between a commensurate and an incommensurate system
cannot be made. We speculate that this indicates a crossover to a depinning
transition with true critical behavior as is reported for sandpile models
or the Fukuyma-Lee-Rice model for charge density waves.

%%%%%%%%%%%%%%%%%%%%%%%%%%%%%%%%%%%%%%%%%%%%%%%%%%%%%%%%%%%%%%%%%%%%%%%

%%%%%%%%%%%%%%%%%%%%%%%%%%%%%%%%%%%%%%%%%%%%%%%%%%%%%%%%%%%%%%%%%%%%%%%
\section{Discussions and Conclusions}
%%%%%%%%%%%%%%%%%%%%%%%%%%%%%%%%%%%%%%%%%%%%%%%%%%%%%%%%%%%%%%%%%%%%%%%

In this work we have developed a model for artificial vortex-flow
channels motivated by recent experiments.\cite{Pruy,Theu,Bess}
We have studied the depinning properties of vortices in channels of
finite length taking into account inhomogeneities such as the sample
boundaries and disorder.

Throughout our analysis, we have neglected the influence of thermal or quantum
fluctuations on the depinning process. To see that this is well-justified with
respect to the experiments that have been performed so far, let us estimate
the magnitude of both types of fluctuations in  conventional
type-II superconductors which have been used to fabricate the channels.
The typical pinning energies of topological lattice defects
are larger than the vortex self energy, which is about $10~eV$,
whereas the thermal energy is  $ \sim 10^{-4} eV$ since the samples are
cooled down to $\lsim 2K$.
Hence, only when the resulting energy barriers are reduced by a factor
$\sim 10^{-5} $ one can expect to observe thermal creep of vortex lattice
defects. Since the barrier energy scales as $ \propto (1-j/j_{c})^{3/2}$, the
width of the current density interval around $ j_{c}$
at which thermal creep can be observed is $ \Delta j \lsim 10^{-3} j_{c} $.

To estimate the relevance of quantum creep, we consider the
effective Euclidean action as a function of the current density,
$ S_{E}^{eff}(j) $. At equilibrium, $j=0$,
$ S_{E}^{eff}(j=0) \approx \xi \hbar^2/(e^2 \rho)(j_{d}/j_{c})^{1/2}$,
see Ref.\ \cite{Blat}. Here the coherence length is $ \xi \sim 10~ nm$,
the resistivity in the normal state
is $ \rho \sim 10^{-6} \Omega m $, the depairing current is
$ j_{d} \sim 10^{9}~ A/m^2 $,
and the critical current is $ j_{c} \sim 10^{6} A/m^2$.
We find $ S_{E}^{eff}(j=0)/\hbar \sim 10^{3} $.
For an overdamped vortex dynamics,
which is characteristic for conventional superconductors,
$  S_{E}^{eff}(j)/\hbar \sim 10^{3} (1-j/j_{c})^{3/4}$. Quantum effects
become irrelevant for $  S_{E}^{eff}(j) \sim \hbar$,
hence $ \Delta j \sim 10^{-4} j_{c} $.
Thus, in artificial vortex-flow channels made of conventional superconductors
thermal and quantum fluctuations only become relevant in an extremely
narrow interval around the critical current which has not been resolved in the
experiments discussed here. In a first step, we have therefore entirely
neglected fluctuations and have studied a purely classical mechanical model
at zero temperature.

First, we have considered
a perfectly ordered vortex lattice in the channel environment.
Starting from a London description of straight vortices in a superconductor, we
evaluate the periodic potential experienced by the vortices in the channel.
The result, which is strictly valid only at
inductions far below $ B_{c2}$ is generalized to higher inductions by
taking into account the vortex core and by
introducing a field dependent coherence length and penetration depth.
We have studied the overdamped dynamics of a chain of interacting vortices
in the effective channel potential. In our simplified description we
have restricted
the considerations to channels with a width of the order of the
lattice spacing, $ w \sim b_{0} $. We have assumed that the resulting
confinement perpendicular to the channel
direction is strong such that the transverse motion of the
vortices in the channel can be neglected. Further, since in all known channel
experiments even for the lowest achievable magnetic inductions
$ \lambda \gsim a_{0} $ holds, to a very good accuracy only the first
harmonic of the periodic potential is kept. These assumptions and restrictions
allow to determine the coefficients of a driven generalized Frenkel-Kontorova
model. For $ \lambda \gsim a_{0} $, where the typical length $l_d$ of a
topological defect in the channel is much larger than the lattice
spacing $a_0$, but still much smaller than the system size $l_s$,
$ l_{s} \gg l_{d} \gg a_{0}$, the dynamics in the channel is
conveniently described by the continuum limit of the Frenkel-Kontorova model
known as the driven sine-Gordon model.

After deriving the coefficients of the Frenkel-Kontorova model, we
investigate the Commensurate-Incommensurate transition commonly known to
occur in this model at thermal equilibrium. Since our focus is on the
depinning problem {\em in the absence of fluctuations}, we
modify the theory. We define a ``mechanical''
Commensurability-Incommensurability transition. It turns out, that the
boundaries of the system play a crucial role if one supposes that the
number of static vortices in the strong pinning environment is constant: the
purely mechanical Commensurate-Incommensurate transition occurs when the entry
barrier for discommensurations at the boundary vanishes.

This concept can be generalized to describe depinning at finite driving
forces. The reason is that the entry barrier for discommensurations is
reduced when a driving force is applied. 
The depinning then occurs at frustrations below the zero force threshold
frustration, at
which the mechanical Commensurate-Incommensurate transition occurs.
In this regime, the entry barrier is by far the largest barrier in the system.

Above the zero-force threshold frustration, in the incommensurate regime,
the entry barrier has vanished and discommensurations enter the system until
the mutual repulsion between the defects prevents new ones to flow in.
The extremely small Peierls-Nabarro barrier
which may arise due to the discreteness of the system \cite{Brau} is not
taken into account.

The depinning scenario depends on both, on the driving force and on the
frustration parameter $ \chi = (a-a_{0})/a_{0}$,
which measures the mismatch between the lattice constant $ a_{0} $ of the
channel environment and the preferred lattice spacing in the channel $ a$.
In the absence of frustration, $ \chi=0 $,
depinning occurs via a trivial homogeneous
solution when the barrier of the tilted washboard potential vanishes and
the threshold force is given by the amplitude of the sinusoidal channel
pinning force, $ f_{c} = \mu $. In the presence of frustration, depinning
occurs via the formation of topological defects at one of the sample
boundaries. For weak frustration, $ |\chi| \ll \chi_{c}(0) $, we
find $ f_{c} = \mu \{ 1- (1/2)[\sqrt{3}~ \chi / \chi_{c}(0)]^{4/3} \} $,
whereas close to the threshold frustration,  $ \chi_{c}(0) -  |\chi|
\ll \chi_{c}(0) $, we obtain
$ f_{c} = (4 \mu / \pi)[1- \chi / \chi_{c}(0)]$
and for $ |\chi| > \chi_{c}(0) $
the depinning force vanishes, $ f_{c} = 0$. By performing a numerical
integration of the equation of motion we have determined the values of
$ f_c(\chi)$ in between these limits.

Further, we have numerically calculated the
force-velocity curves which correspond to the current-voltage characteristics
of the sample. In the commensurate regime, $ |\chi| < \chi_{c}(0) $ we
find a characteristics typical to a saddle-point bifurcation scenario,
$ v \sim (f-f_{c})^{1/2} $, whereas in the incommensurate regime, where
the commensurability gap has vanished, the response is linear, $ v \sim f$.
This behavior is typical for a system with open boundary conditions, where
the density of topological defects is a function of {\em both} the
frustration $\chi$ and the driving force $f$.

In systems with twisted periodic boundary conditions, where the density of
topological defects is constant, the force-velocity characteristics
is different in the incommensurate regime \cite{Bess}.
One observes a linear low-mobility regime for small driving
forces, $ f \ll \mu $. The slope of the linear part is proportional to the
density of topological defects in the system which is {\em fixed} by the
magnitude of the twist at the ends of the system. The slope of the
low-force regime is thus considerably smaller than in the high force regime
$ f \gg \mu $. For $ f \gsim \mu $ the curves show a square-root dependence,
$ v \sim (f-\mu)^{1/2} $ and only in the high force limit they become
linear again. The resulting force-velocity characteristics thus differ
significantly from our results in the incommensurate phase, which is
linear for all forces.

We conclude that the boundary conditions strongly
affect the force-velocity characteristics. The reason is that in frustrated
systems the presence of (open) boundaries supports the formation of
topological defects which
lead to depinning. Further, the entry barriers at the boundaries
determine the rate at which defects enter
the sample, thus ruling the dynamic behavior entirely.
The main problem in determining the behavior of vortex-flow channels is thus
to model realistic sample boundaries. Note that other boundary effects than
considered here might modify the picture. For example, the vortex lattice of
the channel environment may be distorted
due to the presence of screening currents along the sample boundary
causing a local variation of the frustration. Further,
screening currents may lead to Bean-Livingston barriers for vortices
which would increase the energy to form a defect at the sample edge. However,
the conclusion that transport in artificial vortex-flow channels
with a perfectly ordered vortex lattice in the environment
is ruled by the entry barriers at the sample boundary persists even if
further boundary effects are taken into account.

The picture that depinning occurs only via defect formation at the boundaries
does not hold if inhomogeneities are present in the bulk. For example,
local distortions of the vortex lattice in the channel environment caused
by quenched disorder may generate weak spots that support the formation
of vacancy-interstitial pairs at sufficiently large driving force.

To investigate this issue, we have extended the model by accounting for small
static displacements $ {\bf d}({\bf r})$ of the vortices in the channel
environment. We assumed the channel environment to be in the elastic limit,
$ |\nabla \cdot {\bf d}| \sim \xi/R_{c} \ll 1$, where $\xi$ is the coherence
length measuring the typical displacement within a distance given by the
in-plane correlation length $ R_{c} $ for lattice distortions. Further,
we assumed that close to the channel edges $ \lambda \xi /R_{c} \ll a_{0}/2$
in order to obtain a local approximation for the displacement fields of
the static vortices close to the channel edge. Since $ \xi / R_{c} \ll 1 $
this approximation should be valid as long as the penetration depth is
not orders of magnitudes larger than the typical vortex lattice
constants. Further, we took only into account longitudinal displacements along
the channel edge. For narrow channels, $ b_{0} \sim w $, we then obtained a
generalized amplitude and phase-disordered Frenkel-Kontorova model which
in the continuum limit corresponds to a disordered sine-Gordon model.
Transverse displacements imply modifications of the disordered phase and
amplitude of the sinusoidal pinning force and to additional pinning force
terms. This issue and its consequences for depinning have not been considered
here.

In order to gain a basic understanding, we have first investigated depinning
caused by a specific longitudinal vortex-lattice distortion field
along the channel edges. Depending on their location, these distortions cause
additional local frustration of the system modifying the threshold force
for depinning. Distortions at the boundary of the sample affect the
entry barrier for topological defects and cause shifts of the $ \chi - f$
curve along the $\chi$-axis. Local distortions in the bulk are shown to
act as nucleation seeds reducing the threshold force.

Finally, we have studied the effect of small disorder-induced
displacements of vortices in the channel environment.
We model disorder by uncorrelated relative displacements which are
represented by random values that are independent identically distributed
according to a box-distribution.
In the absence of frustration local lattice distortions in the channel
environment lead to an effective channel potential with weak spots. At the
weakest spots vacancy-interstitial pairs are formed when reaching the
depinning threshold force.
In the presence of frustration, a crossover from bulk depinning to boundary
depinning occurs when the entry barrier becomes smaller than the smallest bulk
pinning barrier.
Increasing the frustration the entry barrier is decreased until it becomes
smaller than the typical bulk pinning force for kinks due to
disorder. Applying a finite driving force that is large enough to
overcome the entry barrier, but smaller than the bulk pinning force,
topological defects enter the channel and become pinned in the bulk.
The finite depinning force in this regime is thus again determined by
bulk properties.
Increasing the driving force up to the threshold, topological defects
which travel some distance until becoming trapped are successively introduced.
Above the depinning threshold this leads to a jerky
motion with successive depinning and pinning of topological defects.

We obtained the force-velocity characteristics, the threshold force as
a function of the frustration, and the threshold force as a function
of the disorder for an ensemble of channels with randomly perturbed
channel edges.
Since the theoretical results are
only valid for narrow channels with one mobile row, weak longitudinal
disorder, and since the squeezing effects
of channel edge currents were not taken into account,
the interpretation of the available experimental data remains speculative.
It would thus be interesting to extend the model to include
these additional properties.

Realizing that the Frenkel-Kontorova model and the sine-Gordon model
are equivalent, the question arises, whether the disordered Frenkel-Kontorova model
studied here is  related to the
Fukuyama-Lee-Rice model (FLRM) \cite{Fuku,Lee} for
charge density waves.
This question is especially interesting with respect to the characteristics
of the depinning transition.
The FLRM and simplified versions have been studied both
analytically \cite{Fish,Nara} and numerically,\cite{Copp,Midd,Myer}
mostly in higher dimensions.
The FLRM possesses a
phase-disordered sinusoidal potential where the phases are chosen
randomly from an interval $ [-\pi,\pi]$. This model shows critical behavior
for $d<4$.
Approaching the threshold from below, the critical state is formed by the
release of avalanches characterized by typical sizes that diverge with
a power-law behavior. Above threshold the motion is typically jerky
\cite{Copp} and the velocity shows a power-law behavior
$ v \sim (f-f_{c})^{\nu}$, where $\nu$ depends on the
dimensionality of the system.

One is tempted to say that vortex flow channels provide a physical
realization of the one-dimensional FLRM. However, we find that the
depinning process and the dynamics above threshold strongly depend on the
type of boundaries at the sample edges, the
strength of the frustration, and the strength and type of disorder.
In finite weakly disordered systems
as studied in this work three depinning regimes can be identified.
Increasing the frustration depinning first takes place via defect
nucleation at weak spots in the bulk, then via defect nucleation
at the boundary, and finally by releasing pinned pre-existing
defects when the frustration is so strong that the entry barrier
for defects has become irrelevant.

This is indeed very different compared to the FLRM where system boundaries
are not taken into account and the disorder is of a different type.
If at all, characteristics of the FLRM like the avalanching below depinning
threshold, the roughness of the critical state at threshold, and the jerky
motion above the depinning threshold resemble to what we observe in the
incommensurate state in the presence of weak disorder.

The main difference to the problem treated here is that the FLRM considers
a system close to thermal equilibrium whereas here we are interested in the
depinning behavior far from equilibrium.
We identify typical configurations that act as sources which produce
vacancy-interstitial pairs and thus
lower the depinning threshold. These we call weak spots.
We find also other configurations that  pin
the topological defects. The system is static if all topological defects
that enter the system via the
boundaries or are induced at weak spots are trapped by lattice distortions
of the pinning type. Depinning takes place
when  the density of topological defects
becomes larger than the density of pinning sites. It is thus clear that an
enhancement of the depinning threshold can only occur in
systems with a considerable amount of pinning sites formed by lattice
distortions in the channel.

\section{Acknowledgments}

T.\ D.\ acknowledges financial support from the DFG-Projekt No. MO815/1-1
and from the Graduierteenkolleg
``Physik nanostrukturierter Festk\"orper'', University of Hamburg.
R.\ B.\ and P.\ K.\ are supported by
the Nederlandse Stichting voor Fundamenteel Onderzoek der Materie (FOM)
and C.\ M.\ S.\ is supported by the Swiss National Foundation under
grant 620-62868.00.

\end{document}